\documentclass[12pt]{iopart}

\usepackage{iopams}
\usepackage{graphicx}
\graphicspath{{./}{./plts/}}
\usepackage{color}
\usepackage{subfig}
\usepackage{hyperref}

\begin{document}

\title{Dynamics of a $\mathbb{Z}_2$ symmetric
EdGB gravity in spherical symmetry}

\author{Justin L. Ripley and Frans Pretorius}

\address{
 Department of Physics, Princeton University, Princeton, New Jersey 08544, USA.
}
\ead{jripley@princeton.edu and fpretori@princeton.edu}
\vspace{10pt}
\begin{indented}
\item[]July 2020
\end{indented}

\begin{abstract}
	We report on a numerical investigation of black hole evolution
in an Einstein dilaton Gauss-Bonnet (EdGB)
gravity theory where the Gauss-Bonnet coupling
and scalar (dilaton) field potential are symmetric under a global change in
sign of the scalar field (a ``$\mathbb{Z}_2$'' symmetry).
We find that for sufficiently small Gauss-Bonnet couplings Schwarzschild black
holes are stable to radial scalar field perturbations, and are unstable 
to such perturbations for sufficiently large couplings.
For the latter case, we provide numerical evidence that there is a band of coupling
parameters and black hole masses where the end states are stable scalarized
black hole solutions, in general agreement with the results of
Macedo et. al. \cite{Macedo:2019sem}.
For Gauss-Bonnet couplings larger than those in the stable
band, we find that an ``elliptic region'' forms outside of the black hole horizon,
indicating the theory does not possess a well-posed
initial value formulation in that regime.
\end{abstract}

%
\vspace{2pc}


\section{Introduction}
The recent advent of the field of gravitational wave astronomy has
led to increased interest in testing modifications/extensions of
general relativity (GR) (e.g. \cite{TheLIGOScientific:2016src}).
This requires understanding the dynamics of those
theories during binary black hole inspiral and merger.
While perturbative solutions
to modified gravity theories may be sufficient to describe their predictions
during the inspiral phase, where the black holes are well separated
and gravitational fields are relativity weak, the loudest signal 
during binary inspiral comes from the merger phase, where gravity is
in the strong field, dynamical regime (e.g. \cite{Yunes:2016jcc}).
Order reduction solutions to modified gravity theories
offer a potential route to extracting predictions from those
theories (e.g. \cite{Okounkova:2017yby,Okounkova:2018pql}).
Solving the full equations of motion without using order reduction
potentially offers certain advantages though, as in this approach
one does not have to consider delicate issues
(e.g. ``secular effects'') regarding when
the perturbative order-reduction approximation may fail.
Motivated by this, here we consider the nonlinear spherical
dynamics of a variant of Einstein dilaton Gauss-Bonnet (EdGB) gravity
in spherical symmetry.
While much of the relevant physics to gravitational wave astronomy
(most importantly, gravitational waves) are not present in
spherical symmetry, our study explores the dynamics
of the scalar degree of freedom in the theory,
which will be relevant when solving
for the dynamics of the theory in less symmetrical spacetimes.

	The variant of EdGB gravity we study has been shown to possess
spherically symmetric scalarized
black hole solutions that are stable to linear, radial perturbations
\cite{Macedo:2019sem}.
We find numerical
evidence that stable scalarized black holes can be formed in
this theory, but for large enough Gauss-Bonnet coupling the theory dynamically
loses hyperbolicity. Our results, along with the recent work
of \cite{Kovacs:2020ywu,Kovacs:2020pns},
suggests that for sufficiently small couplings
scalarized black hole solutions in this theory
could be studied in binary inspiral and merger scenarios. 

An outline of the remainder of the paper is as follows.
In Sec.\ref{sec:eom} we
describe the particular EdGB theory we consider,
and the resulting equations of motion in spherical symmetry.
In Sec.\ref{sec:earlier} we briefly summarize earlier
work on scalarized black holes in this theory.
In Sec.\ref{sec:description_of_code_and_simulations} we describe 
our numerical code.
In Secs. \ref{sec:id} and \ref{sec:res} we describe our initial data and
evolution results respectively.
We present concluding remarks in Sec.~\ref{sec:discussion}.
In \ref{sec:comparison} we describe the characteristic analysis
we employ during evolution, contrasting it with  linear perturbation analysis,
and in \ref{sec:approx_scalarized_profile} we derive the approximate
scalarized initial data we used in some of our simulations.

	We use geometric units ($8\pi G=1$, $c=1$)
and follow the conventions of
Misner, Thorne, and Wheeler \cite{misner1973gravitation}. 
\section{Equations of motion}
\label{sec:eom}
	The action for the class of EdGB gravity theories we consider is 
\begin{equation}
\label{eq:EdGBAction}
	S = \int d^4x\sqrt{-g}
	\left(
		\frac{1}{2}R 
	- 	\frac{1}{2}(\nabla\phi)^2 
	- 	V(\phi) 
	+ 	W(\phi)\mathcal{G}
	\right)
	,
\end{equation}
	where $\phi$ is a scalar field (the ``dilaton'' field),
$g$ is the metric determinant,
$V(\phi)$ and $W(\phi)$ are (so far unspecified) functions of $\phi$,
$R$ is the Ricci scalar, and $\mathcal{G}$ is the Gauss-Bonnet scalar:
\begin{equation}
	\mathcal{G}
	\equiv
	\frac{1}{4}\delta^{\mu\nu\alpha\beta}_{\rho\sigma\gamma\delta}
	R^{\rho\sigma}{}_{\mu\nu}
	R^{\gamma\delta}{}_{\alpha\beta}
	,
\end{equation}
	where $\delta^{\mu\nu\alpha\beta}_{\rho\sigma\gamma\delta}$
is the generalized Kronecker delta tensor and
$R^{\rho\sigma}{}_{\mu\nu}$ is the Riemann tensor.
Varying \ref{eq:EdGBAction} with respect to the metric and
scalar fields, the EdGB equations of motion are 
\numparts
\begin{eqnarray}
\label{eqns:einstein_eqns}
\label{eq:tensor_eom}
	E^{(g)}_{\mu\nu}&\equiv&
	R_{\mu\nu} - \frac{1}{2}g_{\mu\nu}R
+ 	2\delta^{\gamma\delta\kappa\lambda}_{\alpha\beta\rho\sigma}
	R^{\rho\sigma}{}_{\kappa\lambda}
	\left(\nabla^{\alpha}\nabla_{\gamma}W(\phi)\right)
	\delta^{\beta}{}_{(\mu}g_{\nu)\delta}
	- T_{\mu\nu} 
	= 0 
	, \\
	T_{\mu\nu}
 	&\equiv&
	\nabla_{\mu}\phi\nabla_{\nu}\phi
-	g_{\mu\nu}\left[\frac{1}{2}(\nabla\phi)^2+V(\phi)\right]
	, \nonumber \\
\label{eq:scalar_eom}
	E^{(\phi)}&\equiv&
		\nabla_{\mu}\nabla^{\mu}\phi 
	-	V^{\prime}(\phi)
	+ 	W^{\prime}(\phi)\mathcal{G} 
	= 0
	,
\end{eqnarray}
\endnumparts
	where the ``prime'' operation defines a derivative
with respect to the scalar field $\phi$;
e.g. $V^{\prime}(\phi)\equiv dV(\phi)/d\phi$.
We see that if $\phi=const.$, then the Gauss-Bonnet scalar
term does not contribute to the equations of motion;
this is a consequence of the fact that the
Gauss-Bonnet scalar is locally a total derivative
in four dimensional spacetime  (see e.g. \cite{nakahara2018geometry}).
In this article we will solve Eqs.~\ref{eqns:einstein_eqns}
with the following choice for functions $V(\phi)$ and $W(\phi)$
\numparts
\begin{eqnarray}
\label{eq:general_V}
	V(\phi)
	= &
	\frac{1}{2}\mu^2\phi^2
+	\lambda\phi^4
	, \\
\label{eq:general_W}
	W(\phi)
	= &
	\frac{1}{8}\eta\phi^2
	,
\end{eqnarray} 
\endnumparts
where $\mu,\lambda$ and $\eta$ are constant parameters\footnote{These
are the same potentials used by the authors in \cite{Macedo:2019sem};
for the particular numerical values of the coupling parameters
$\mu$, $\lambda$, and $\eta$ we have rescaled ours to match 
those in \cite{Macedo:2019sem}, taking
into account our different choice of scalar field normalization.}.
In our earlier works on EDGB gravity we used the
potentials $V=0,W=\lambda\phi$~\cite{Ripley:2019hxt,Ripley:2019aqj,Ripley:2019irj});
that choice encompasses the leading order
term for shift symmetric $\phi\to\phi+const.$ EdGB gravity models. 
Here, we focus on the form \ref{eq:general_V} and \ref{eq:general_W},
the leading order contributions to theories
that have $\phi\to-\phi$ ($\mathbb{Z}_2$) symmetry
\cite{Macedo:2019sem}\footnote{We note though that we do not
have a $(\nabla\phi)^4$ term in our action,
even though it is symmetric under $\phi\to-\phi$ and
is of no higher order than the Gauss-Bonnet term, as in this work we
only wish to consider how black hole dynamics are affected by the addition
of a scalar Gauss-Bonnet coupling.}.
We note that with the code we have developed we could in principle
investigate the spherical dynamics of EdGB gravity theories with
arbitrary functions $V(\phi)$ and $W(\phi)$ (see Sec.~\ref{sec:code}). 

	The potential $W(\phi)$ satisfies
an ``existence condition''for pure GR solutions \cite{Silva:2017uqg}:
at the minimum $\phi_0$ of the potential $W(\phi)$, here $\phi_0=0$,
the potential satisfies:
\begin{equation}
\label{eq:existence_condition_GR_W}
	W^{\prime}\big|_{\phi=0}=0
	.
\end{equation} 
	That is, for $W$ that obey Eq.~\ref{eq:existence_condition_GR_W}
we can obtain dynamical GR solutions with $\phi=0$,
as the $\phi$ will not be sourced by curvature
(although it could still potentially be sourced by other matter fields,
and once there are regions where $\phi\neq0$, $W^{\prime}\neq0$
and the scalar field could then be sourced by curvature terms).

	We evolve this system in Painlev\'{e}-Gullstrand (PG)-like coordinates 
        (e.g. \cite{Adler:2005vn,Ziprick:2008cy,Kanai:2010ae,Ripley:2019tzx})
\begin{equation}
	ds^2
	=
-	\alpha(t,r)^2dt^2
+	\left(dr+\alpha(t,r)\zeta(t,r)dt\right)^2
+	r^2\left(
		d\vartheta^2
	+	\mathrm{sin}^2\vartheta d\varphi^2
	\right)
	,
\end{equation} 
so-named since $t={\rm const.}$ cross sections are spatially flat 
as in the PG coordinate representation of the Schwarzschild black hole
(which in these coordinates is given by $\alpha=1,\zeta=\sqrt{2 m/r}$).
	
	We define the variables
\numparts
\begin{eqnarray}
\label{eqns:defs_PQ}
\label{eq:def_Q}
	Q
	\equiv &
	\partial_r\phi
	, \\
\label{eq:def_P}
	P
	\equiv &
	\frac{1}{\alpha}\partial_t\phi
-	\zeta Q	
	=
	n^{\mu}\partial_{\mu}\phi
	,
\end{eqnarray}
\endnumparts
	and obtain the following system of evolution
equations for $\phi$, $Q$ and $P$:
\numparts
\begin{eqnarray}
\label{eq:evolution_phi}
	E_{(\phi)}
	\equiv &
		\partial_t\phi
	-	\alpha\left(P+\zeta Q\right)
	=
	0
	, \\	
\label{eq:evolution_Q}
	E_{(Q)}
	\equiv &
	\partial_tQ
-	\partial_r\left(
		\alpha\left[
			P
		+	\zeta Q
		\right]
	\right)
	=
	0
	, \\
\label{eq:evolution_P}
	E_{(P)}
	\equiv &
	\mathcal{A}_{(P)}\partial_tP
+	\mathcal{F}_{(P)}
	=
	0
	. 
\end{eqnarray}
\endnumparts
	The evolution equation for $\phi$
(Eq.~\ref{eq:evolution_phi}) follows
from the definition of $P$ (Eq.~\ref{eq:def_P}), and
the evolution equation for $Q$ (Eq.~\ref{eq:evolution_Q})
follows from taking the radial derivative of Eq.~\ref{eq:evolution_phi}.
The evolution equation for $P$ (Eq.~\ref{eq:evolution_P})
comes from taking algebraic combinations of Eq.~\ref{eq:scalar_eom} and
the $tr$, $rr$, and $\vartheta\vartheta$ components of
Eq.~\ref{eq:tensor_eom}
(c.f. \cite{Ripley:2019irj}).
The quantities  $\mathcal{A}_{(P)}$ and $\mathcal{F}_{(P)}$ are lengthy
expressions of $\{\alpha,\zeta,P,Q\}$ and their radial derivatives.
In the limit $W=0$, Eq.~\ref{eq:evolution_P} reduces to
\begin{equation}
	\partial_tP
-	\frac{1}{r^2}\partial_r\left(r^2\alpha\left[Q+\zeta P\right]\right)
+	\alpha V^{\prime}
	=
	0
	.
\end{equation} 
Interestingly, in PG coordinates the Hamiltonian and momentum constraints do not
change their character as elliptic differential equations
(or more properly, as ordinary differential equations in the radial coordinate $r$)
going from GR to EdGB gravity:  
\numparts
\begin{eqnarray}
\label{eq:Hamiltonian_constraint}
\fl	E^{(g)}_{\mu\nu}n^{\mu}n^{\nu}
	\propto& 
	\left(
		1
	-	\frac{8QW^{\prime}}{r}
	-	\frac{12P\zeta W^{\prime}}{r}
	\right)\partial_r\zeta
+	\left(
		\zeta
	-	\frac{8Q\zeta W^{\prime}}{r}
	-	\frac{12P\zeta^2W^{\prime}}{r}
	\right)\frac{\partial_r\alpha}{\alpha}
	\nonumber \\ & 
+	\frac{\zeta}{2r}
-	\frac{r}{2\zeta}\rho
-	\frac{4W^{\prime}\zeta }{r}\partial_rQ
-	\frac{4W^{\prime\prime}Q^2 \zeta}{r}
	= 0
	, \\
\label{eq:momentum_constraint}
\fl	E^{(g)}_{\mu r}n^{\mu}
	\propto& 
	\left(
		1
	-	\frac{8W^{\prime}Q}{r}
	-	\frac{8W^{\prime}P\zeta}{r}
	+	\frac{4W^{\prime}Q\zeta^2}{r}
	\right)\frac{\partial_r\alpha}{\alpha}
-	\frac{r}{2\zeta} j_r
	\nonumber \\ &
+	\frac{4W^{\prime}\zeta}{r}\partial_rP
+	\frac{4W^{\prime}Q\zeta}{r}\partial_r\zeta
+	\frac{4W^{\prime\prime}PQ\zeta}{r}
	= 0
	,
\end{eqnarray}
\endnumparts
	where 
\numparts
\begin{eqnarray}
\label{eqns:defs_source_terms}
\label{eq:def_rho}
	\rho
	&\equiv&
	n^{\mu}n^{\nu}T_{\mu\nu}
	=
	\frac{1}{2}\left(P^2+Q^2\right) + V
	, \\
\label{eq:def_jr}
	j_{\kappa}
	&\equiv&
-	\gamma_{\kappa}{}^{\mu}n^{\nu}T_{\mu\nu}
	=
	-PQ
	,
\end{eqnarray}
\endnumparts
	and $n_{\mu}\equiv(-\alpha,0,0,0)$. 
\section{Earlier work on $\mathbb{Z}_2$ symmetric EdGB gravity}\label{sec:earlier}
	Here we briefly review recent work on $\mathbb{Z}_2$ symmetric EdGB
gravity theories. 
	Potentials of the form $V=0$ and $W=w_2\phi^2$ were considered
in \cite{Silva:2017uqg}. There the authors found scalarized black hole
solutions, which were subsequently found to be mode unstable
to radial perturbations in \cite{Blazquez-Salcedo:2018jnn}.
Subsequently, radial mode-stable scalarized black hole solutions
were found for couplings of the form
$V=0$, $W=w_2\phi^2+w_4\phi^4$ in 
\cite{PhysRevD.99.044017,PhysRevD.99.064011}. 
Couplings of the form $V=0$ and $W=c_0+\mathrm{exp}\left(c_e\phi^2\right)$
have been investigated in
\cite{PhysRevLett.120.131103,Blazquez-Salcedo:2018jnn}, and also
give rise to radially stable scalarized black hole solutions.
	
	The authors in \cite{Macedo:2019sem} introduced the model we 
study in this article, $V=\mu^2\phi^2+2\lambda\phi^4$, $W=\eta\phi^2/4$.
This model is motivated by effective field theory arguments: assuming
the action is invariant under the $\mathbb{Z}_2$ symmetry $\phi\to-\phi$,
the action contains all terms (except for the term $(\nabla\phi)^4$,
which we do not consider in this article)
of mass dimension equal to or less than the mass
dimension of $\phi^2\mathcal{G}$, which is the lowest order term that
couples $\phi$ to the Gauss-Bonnet scalar subject to this symmetry.
Though like the authors in \cite{Macedo:2019sem},
while we motivate this model from effective field theory we in fact
will treat the theory as a complete classical field theory, and 
consider exact (to within numerical truncation error)
solutions to the equations of motion.
	
	In \cite{Macedo:2019sem}, the authors found
scalarized black hole solutions stable to linear radial perturbations
for certain ranges of the dimensionless parameters
\numparts
\begin{eqnarray}
\label{eq:hat_M}
	\hat{M}
	\equiv&
	\frac{M}{\eta^{1/2}}
	,\\
\label{eq:hat_Phi}
	\hat{\Phi}
	\equiv&
	\frac{\Phi}{\eta^{1/2}}
	,\\
\label{eq:hat_mu}
	\hat{\mu}
	\equiv&
	\mu\eta^{1/2}
	,\\
\label{eq:hat_lambda}
	\hat{\lambda}
	\equiv&
	\lambda\eta
	,
\end{eqnarray}
\endnumparts
	where $M$ is the asymptotic
mass of the black hole plus scalar field configuration
and $\Phi$
is the asymptotic scalar ``charge'',
i.e. for scalarized black hole solutions to the theory
it is the coefficient to the leading nonzero
term in $\phi(t,r)$ as $r$ goes to infinity \cite{Macedo:2019sem} 
\begin{equation}
	\lim_{r\to\infty}\phi(t,r)
	\sim
	e^{-\mu r}\left(\frac{1}{r}\Phi+\cdots\right)
	.
\end{equation}

A nonzero value of $\hat{\lambda}$ is
necessary to have radially linear mode stable scalarized black hole solutions,
and the value of $\hat{\lambda}$ needed to stabilize a given scalarized
black hole increases as $\hat{\mu}$ increases \cite{Macedo:2019sem}.
For example, when $\hat{\mu}=0$,
the minimum value of $\hat{\lambda}\approx0.2$.
There is also a maximum value of $\hat{M}$ for a scalarized black hole
(above this the Schwarzschild solution is stable); for example
for $\hat{\mu}=0$, 
found this maximum to be $\hat{M}\approx0.6$.
\section{Description of code and simulations}
\label{sec:description_of_code_and_simulations}
\subsection{Code description}
\label{sec:code}
	Our basic evolution strategy is the same as in
\cite{Ripley:2019aqj}: we freely evolve $P$ and $Q$
using Eqs.~\ref{eq:evolution_Q} and \ref{eq:evolution_P}, and solve
for $\alpha$ and $\zeta$ using the constraint equations,
Eqs.~\ref{eq:Hamiltonian_constraint} and \ref{eq:momentum_constraint}.
The boundary condition for $\zeta$ at the excision boundary is
obtained by freely evolving it using the $E_{tr}$ equation of motion
(with algebraic combinations of the other equations of motion to remove time
derivatives of $\alpha$ and $P$). We do not need to impose
any boundary condition for $\alpha$ on the excision boundary
due to the $\alpha\to\alpha+c(t)$
residual gauge symmetry in PG coordinates. The full equations of motion
for $P$ and the freely evolved $\zeta$ on the boundary
are long and unenlightening, although see Appendix C of \cite{Ripley:2019aqj}
for their full form in PG coordinates for the special case $V=0$ (and note
that there we use the notation $W\equiv f$).

As in \cite{Ripley:2019aqj}, we solve the constraint equations using the
trapezoid rule (a second order method) with relaxation.
Unlike \cite{Ripley:2019aqj} though,
we solve the evolution equations for $\{P,Q\}$, and
$\zeta$ on the excision boundary using a fourth order method of lines
technique: we use fourth order centered difference stencils to evaluate the
spatial derivatives, and
evolve in time with a fourth order Runge-Kutta integrator
(e.g. \cite{Numerical_recipes,gustafsson1995time}). At the excision boundary
and the boundary at spatial infinity we used one-sided difference stencils
to evaluate spatial derivatives.
We solve the equations over a single grid (unigrid evolution).
An example of a convergence study with the
independent residual $E_{rr}$ is shown in
Fig.~\ref{fig:indep_res_ellp_formation}.  

	The code is written in C++, and can be accessed at 
\cite{justin_ripley_2020_3873503}.
\subsection{Diagnostics}
	Our code diagnostics are as in
\cite{Ripley:2019hxt,Ripley:2019aqj,Ripley:2019irj},
which we very briefly review here. 
The mass of a given simulation is determined by the value of the
Misner-Sharp mass at spatial infinity:
\begin{equation}
	m_{total}
	=
	\lim_{r\to\infty}
	m_{MS}(t,r)
	=
	\lim_{r\to\infty}
	\frac{r}{2}\zeta(t,r)^2
	.
\end{equation}
	Another diagnostic we compute is the radial characteristic speed of
the dynamical degree of freedom of EdGB gravity in spherical symmetry.
To compute the radial characteristic speeds,
we compute the characteristic vector $\xi_a$ by finding the zeros
of the characteristic equation: 
\begin{eqnarray}
\label{eq:characteristic_eqn}
	\mathrm{det}\left[
		\pmatrix{
			\delta E_{(P)}/\delta(\partial_aP)
		&	\delta E_{(P)}/\delta(\partial_aQ)
		\cr	\delta E_{(Q)}/\delta(\partial_aP)
		&	\delta E_{(Q)}/\delta(\partial_aQ)
		}
		\xi_a
	\right]
	=
	0
	.
\end{eqnarray}
	and then compute the ingoing and outgoing characteristic
speeds $c_{\pm}\equiv\mp\xi_t/\xi_r$.
Expanding Eq.~\ref{eq:characteristic_eqn} gives us a quadratic equation
for the radial characteristic speeds
\begin{equation}
\label{eq:cartoon_characteristics}
	\mathcal{A}c^2+\mathcal{B}c+\mathcal{C}=0
	,
\end{equation}
	where $\mathcal{A}$, $\mathcal{B}$, and $\mathcal{C}$ are
complicated functions of the metric and scalar fields, and their
time and radial derivatives (c.f. \cite{Ripley:2019aqj}).
The discriminant
$\mathcal{D}\equiv \mathcal{B}^2-4\mathcal{A}\mathcal{C}$
determines the hyperbolicity of the theory at each spacetime point.
Where $\mathcal{D}>0$, the radial characteristic speeds are
real and the equations of motion for the theory at that spacetime
point are hyperbolic.
Where $\mathcal{D}<0$, the radial characteristic speeds are imaginary
and the equations of motion for the system defined by
Eq~\ref{eq:evolution_phi}, Eq~\ref{eq:evolution_Q},
and Eq~\ref{eq:evolution_P} at that spacetime point
are elliptic \footnote{Note that even though
our evolution system consists of three transport equations, the system is
degenerate and describes the evolution for a single degree of freedom,
which we can define the ingoing and outgoing radial characteristic for.
The system defined by
Eq~\ref{eq:evolution_phi}, Eq~\ref{eq:evolution_Q}, and Eq~\ref{eq:evolution_P}
is degenerate
as $Q\equiv\partial_r\phi$, and the equation of motion for $Q$ is just the
radial derivative of the equation of motion for $\phi$.}. 
We discuss the difference between computing the radial characteristics
and computing the modes of radial linear perturbations about a stationary
solution in \ref{sec:comparison}.

	We determine the location of the black hole horizon in our simulation
by computing the location of the marginally outer trapped surface (MOTS) 
for outgoing null characteristics \cite{Thornburg2007}.
In PG coordinates the radial null characteristics are
\begin{equation}
	c^{(n)}_{\pm}=\alpha\left(\pm1-\zeta\right)
	,
\end{equation} 
	thus the location of the MOTS is at the point $\zeta(t,r)=1$. 
\section{Initial data}\label{sec:id}
\subsection{General considerations}	
	We may freely specify $\phi$ and $\partial_t\phi$
(or, $\phi$ and $P$) on the initial
data surface as $\phi$ satisfies a second order in time wave-like equation,
provided the characteristics for that equation are real.
The variables $\alpha$ and $\zeta$ must satisfy the constraint
equations, and are not freely specifiable.  

	Note that $\phi=0,\partial_t\phi=0$ is not just a solution to the
initial value problem for EdGB gravity with this coupling potential
(\ref{eq:general_V} and \ref{eq:general_W}),
it is also a consistent solution to the full evolution equations
(in fact this
was one criterion used by the authors in \cite{Silva:2017uqg} in constructing
this model of EdGB gravity).
This is in contrast to EdGB gravity with a linear shift-symmetric potential,
where $\phi=0,\partial_t\phi=0$ can by imposed as an initial condition, but then
$\phi$ will generically evolve to non-zero values with time.
\subsection{Black hole initial data with a small exterior scalar pulse}
\label{sec:id_bh_scalar_pulse}
	Our class of initial data is similar to that we used in
\cite{Ripley:2019aqj}: the free initial data is $\phi$ and $P$, from
which we can determine $Q,\alpha,\zeta$. To have black hole initial
data, at the initial excision boundary $x_{exc}$ we set
\begin{eqnarray}
\label{eq:initial_excision_value_al_ze}
	\alpha|_{x=x_{exc}}=&1,\\
	\zeta|_{x=x_{exc}}=&\sqrt{\frac{2M}{r}}
	,
\end{eqnarray} 
	with $\phi$ and $P$ compactly supported away from the
excision boundary, and then integrate outwards in $r$ to obtain
$\alpha$ and $\zeta$ over the initial data surface.
The quantity $M$ in Eq.~\ref{eq:initial_excision_value_al_ze} is the
initial mass of the black hole.

	We choose the following for the exterior scalar pulse 

\numparts
\begin{eqnarray}
\label{eq:bump_id}
\fl	\phi_b(t,r)\big|_{t=0}
	= &
	\cases{
		\frac{a_0}{n_0} \left(r-r_l\right)^2\left(r_u-r\right)^2
		\mathrm{exp}\left[-\frac{1}{r-r_l}-\frac{1}{r_u-r}\right]
		\qquad r_l<r<r_u
		\\
		0
		\qquad
		\mathrm{otherwise}
	}
	, \\
\fl	Q_b(t,r)\big|_{t=0}
	= &
	\partial_r\phi(t,r)\big|_{t=0}
	, \\
\fl	P_b(t,r)\big|_{t=0}
	= &
	0
	,
\end{eqnarray}
\endnumparts
	where $r_u>r_c>r_l>2M$, and the normalization
$n_0$ is chosen such that
\begin{equation}
	\max\phi_b(t,r)\Big|_{t=0}=a_0
	.
\end{equation}
\subsection{Approximate scalarized profile}
\label{sec:initial_data_approx_scalarized}
	We also consider initial data that approximates the
static decoupled scalarized profile
on a Schwarzschild black hole background
(see also \cite{Macedo:2019sem}):
\begin{eqnarray}
\label{eq:scalarized_id}
	\phi(t,r)\big|_{t=0}
	=&
	\frac{\Phi_0}{r}\mathrm{exp}\left(-\mu\left(r-3M\right)\right)
	,\\
	Q(t,r)\big|_{t=0}
	=&
	\partial_r\phi(t,r)\big|_{t=0}
	,
\end{eqnarray}
	where $\Phi_0$ is a constant;
see Appendix \ref{sec:approx_scalarized_profile} for a derivation.
We choose $P$ so that
$\partial_t\phi\approx0|_{t=0}$ (see \ref{eq:def_P}): we set
\begin{eqnarray}
	P(t,r)\big|_{t=0}
	=&
-	Q(t,r)\zeta(t,r)\big|_{t=0}
	,
\end{eqnarray}
	and initially set $\alpha=1$, $\zeta=\sqrt{2M/r}$. We
then resolve the constraints for $\alpha$ and $\zeta$, set $P$ as
above, and iterate this
process until the maximum change in $P$, $|\Delta P|_{\infty}$,
between iterations is less than $10^{-3}$.
\section{Numerical results}\label{sec:res}
	We will only present results for $\eta>0$, and absorb $\eta$ into the
definition of the dimensionless parameters
(Eqs.~\ref{eq:hat_M}-\ref{eq:hat_lambda}).
As one check of the code we did perform
simulations with the GR case $\eta=0$; in these simulations
some of the scalar field would fall into the black hole,
and some of the field would disperse to infinity, leaving a vacuum
Schwarzschild solution behind. This is consistent with
the no hair theorems for a canonically
coupled scalar field with a potential in asymptotically
flat spacetimes (for a review and references
see e.g. \cite{Sotiriou:2015pka}), along with perturbative solutions
of scalar fields around a Schwarzschild black hole background (for
a review and references see e.g. \cite{Berti:2009kk}). 

We find that for both classes of initial data described in the previous section,
we can separate the solutions
into three types: (1) the scalar field disperses, leaving behind
a Schwarzschild black hole, (2) the black hole
scalarizes and approaches a static solution, or
(3) an elliptic region eventually forms outside the black hole
(i.e. exterior to the apparent horizon). 

\subsection{Compact scalar pulse initial data}
\label{sec:compact_scalar_pulse_ID}
\begin{figure*}
	\centering
	\subfloat{{\includegraphics[width=0.7\textwidth]{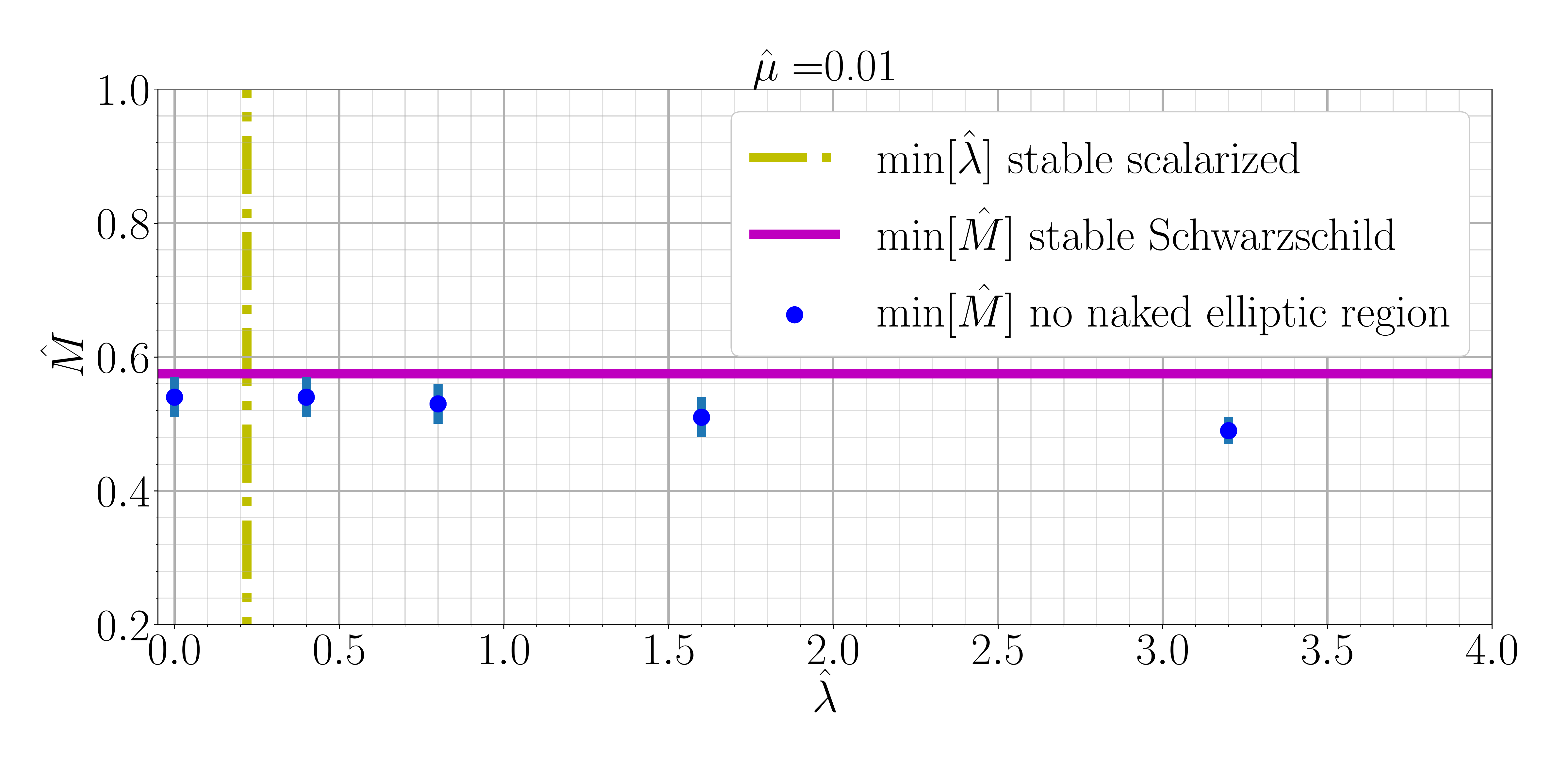}}}%
	\label{subfig:muhat0pt01_compact_scalar}
	\hfill
	\centering
	\subfloat{{\includegraphics[width=0.7\textwidth]{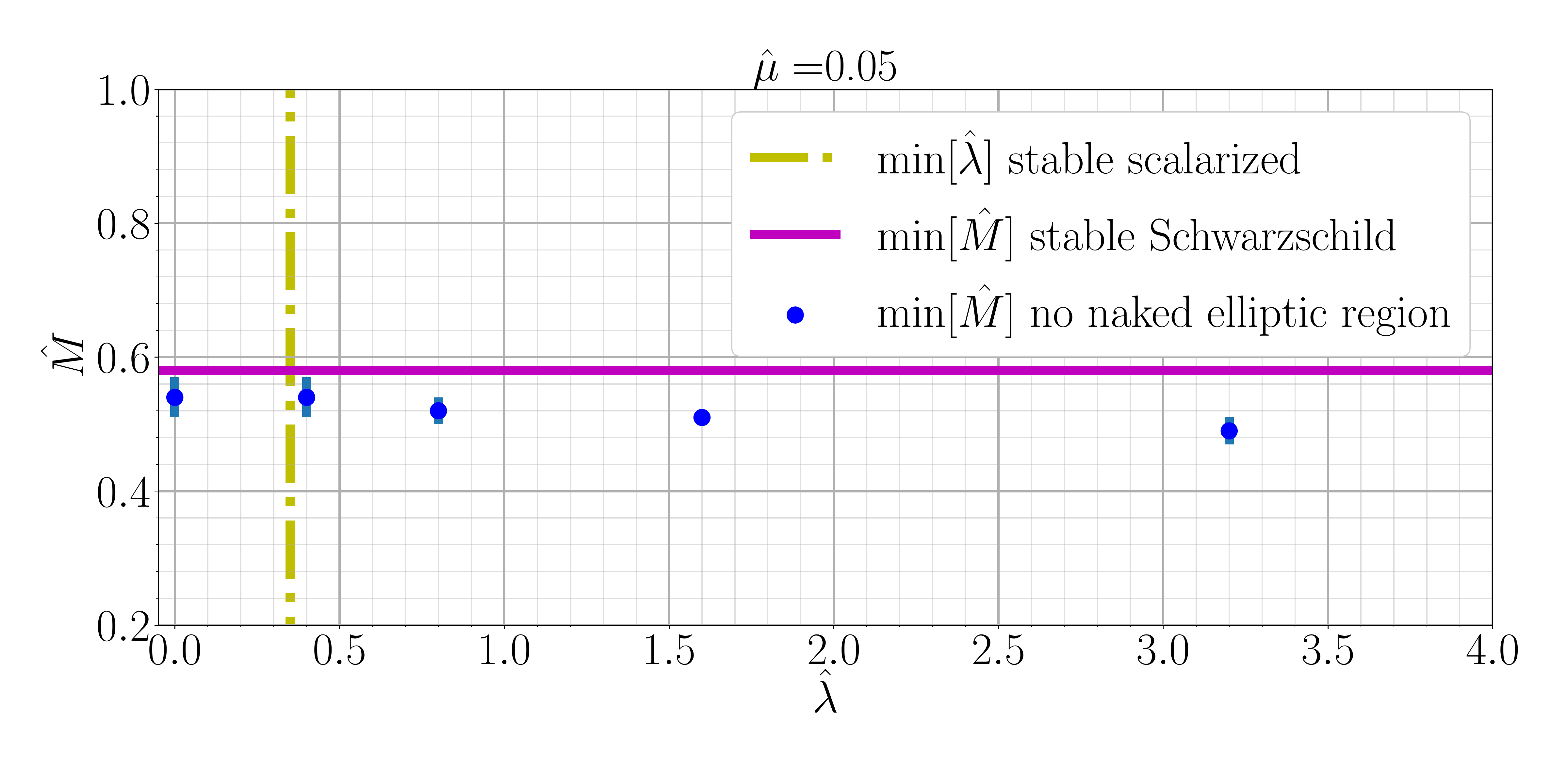}}}%
	\label{subfig:muhat0pt05_compact_scalar}
	\hfill
	\centering
	\subfloat{{\includegraphics[width=0.7\textwidth]{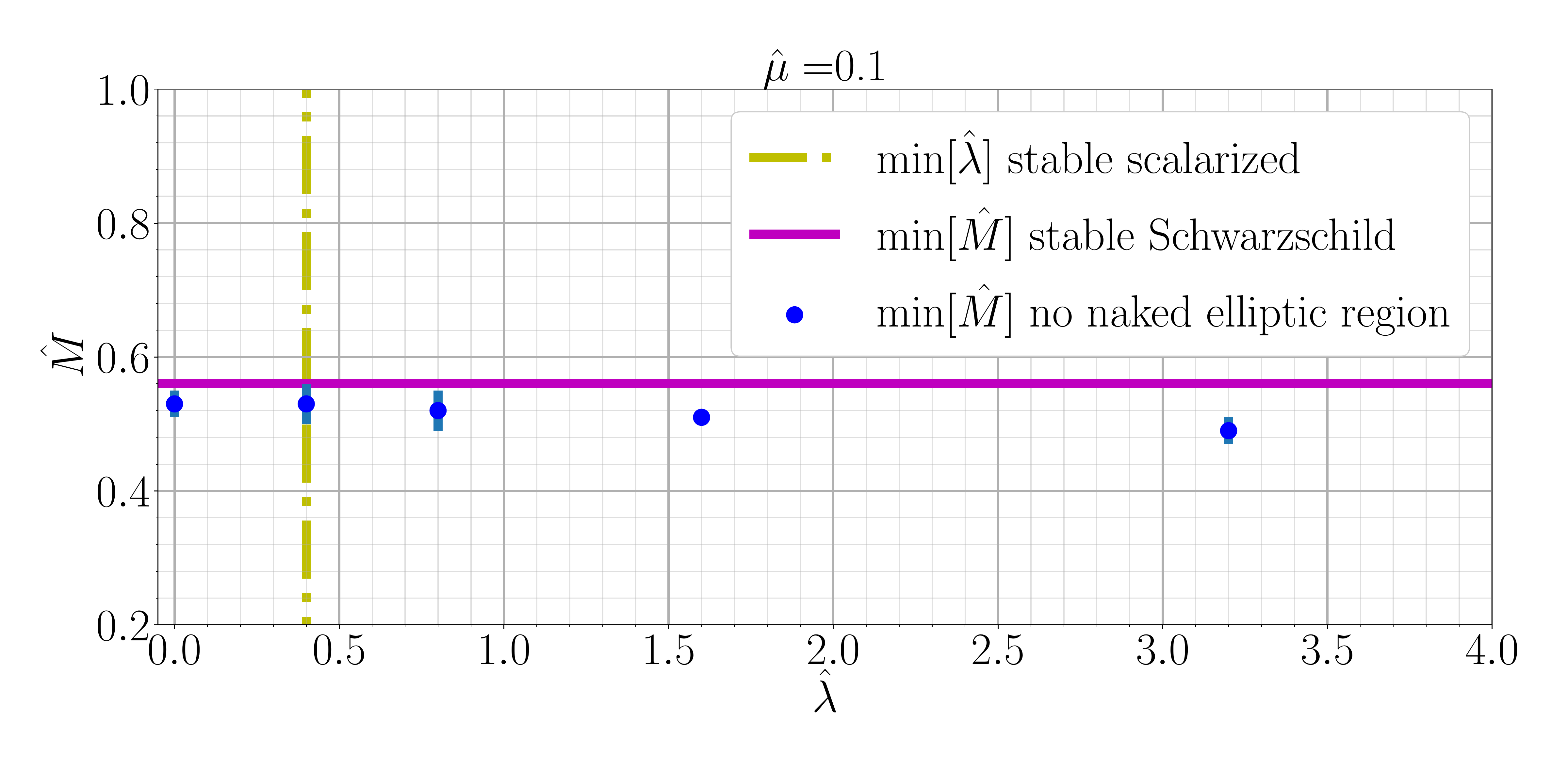}}}%
	\label{subfig:muhat0pt10_compact_scalar}
	\hfill
\caption{
Onset of elliptic region formation, from evolution of perturbed Schwarzschild
initial data as described in Sec.~\ref{sec:compact_scalar_pulse_ID}.
The blue dots are our numerically computed values of 
$\hat{M}$, Eq.~\ref{eq:hat_M},
(for a given $\hat{\lambda}$, Eq.~\ref{eq:hat_lambda}) 
below which an elliptic region eventually forms outside the horizon.
The error bars about each point come from truncation error estimates,
computed by taking the difference of the elliptic onset
point computed with two different resolutions:
$N_x=2^{10}+1$ and $N_x=2^{11}+1$ radial points.
The solid purple horizon line is
the minimum $\hat{M}$ for a stable \emph{Schwarzschild} black hole,
and the dash-dotted yellow vertical line
is the minimum $\hat{\lambda}$
for a stable \emph{scalarized} black hole, with respect to linear
radial perturbations according to the analysis of \cite{Macedo:2019sem}.
}
\label{fig:compact_scalar}
\end{figure*}

	In Fig.~\ref{fig:compact_scalar} we show several points
on a plot of $\hat{M}$ versus $\hat{\lambda}$
(for the definition of $\hat{M}$ and $\hat{\lambda}$
see respectively Eq.~\ref{eq:hat_M} and Eq.~\ref{eq:hat_lambda}),
indicating the division between evolution that
forms elliptic regions and that which does not, beginning from the
compact pulse initial data described in Sec.~\ref{sec:id_bh_scalar_pulse}.
To arrive at the numerical data points in the figure, we fixed the initial data
parameters $r_l$, $r_u$, and $a_0$ and black hole mass. Specifically, we chose
$r_l=0.24$, $r_u=32$, $a_0=5\times10^{-3}$ (~\ref{eq:bump_id}), 
and initial Schwarzschild black hole mass $M\approx10$
(\ref{eq:initial_excision_value_al_ze}).
The contribution of the scalar field to the total mass of the spacetime
is $M_{\phi}\sim9.6\times10^{-3}$ \footnote{We estimate the scalar field
contribution to the total asymptotic Misner-Sharp mass
by subtracting off from the initial Misner-Sharp mass
the mass of the black hole initial data we had before
we resolved the constraints with scalar field initial data.}.
We then performed a bisection search for the value of $\eta$ that would
lead to regular evolution within a run time of $t/M\sim200$ (sufficiently
long for the solution to settle to a near-stationary state
if no elliptic region formed).
As we varied $\eta$, we varied $\mu$ and $\lambda$ such that $\hat{\mu}$
and $\hat{\lambda}$ remained
fixed (see Eqs.~\ref{eq:hat_M}-\ref{eq:hat_lambda}).
Fig.~\ref{fig:compact_scalar} shows that the theory can remain hyperbolic
for perturbations of at least some black hole solutions. 

Also overlaid on Fig.~\ref{fig:compact_scalar}
are the estimated minimum $\hat{\lambda}$, $\mathrm{min}\hat{\lambda}$,
to have a stable scalarized black hole solution, and the
minimum $\hat{M}$, $\mathrm{min}\hat{M}$,
to have a stable Schwarzschild black hole
solution with respect to linear radial perturbations computed in
\cite{Macedo:2019sem}.
From those results one would expect for $\hat{M}<\mathrm{min}\hat{M}$
and for $\hat{\lambda}>\mathrm{min}\hat{\lambda}$,
Schwarzschild black holes would be unstable under radial scalar field
perturbations to forming (stable) scalarized black hole solutions.
Our numerical results are in general agreement with this reasoning (except
that for small enough $\hat{M}$ a elliptic region grows outside of the
black hole horizon).
In particular, above the horizontal purple line in Fig.~\ref{fig:compact_scalar}
initial small perturbations in $\phi$ decay and leave behind
a Schwarzschild black hole, while below this and to the right of the dashed
yellow line the black holes scalarize (and the solutions above the line
implied by the blue dots are free of elliptic regions exterior to the
black hole horizon).
For $\hat{\lambda}$ to the left of the yellow dashed line, we find
only two regimes: either the scalar field disperses and the solution settles
to a Schwarzschild black hole solution, or a naked elliptic region forms. 
From Fig.~\ref{fig:compact_scalar}, we see that the dividing
line between elliptic region formation and Schwarzschild end state solutions
for $\hat{\lambda}<\mathrm{min}(\hat{\lambda})$
does not quite lie on the value of $\min(\hat{M})$ predicted for
stable Schwarzschild black hole solutions by \cite{Macedo:2019sem}.
That being said, our value is not in significant ``tension'' with what they
computed, given the estimated errors of our calculation and given
that the authors in \cite{Macedo:2019sem} performed a linear perturbation
analysis about a static background.

\begin{figure*}
	\centering
	\subfloat[Scalar field disperses]{{\includegraphics[width=0.7\textwidth]{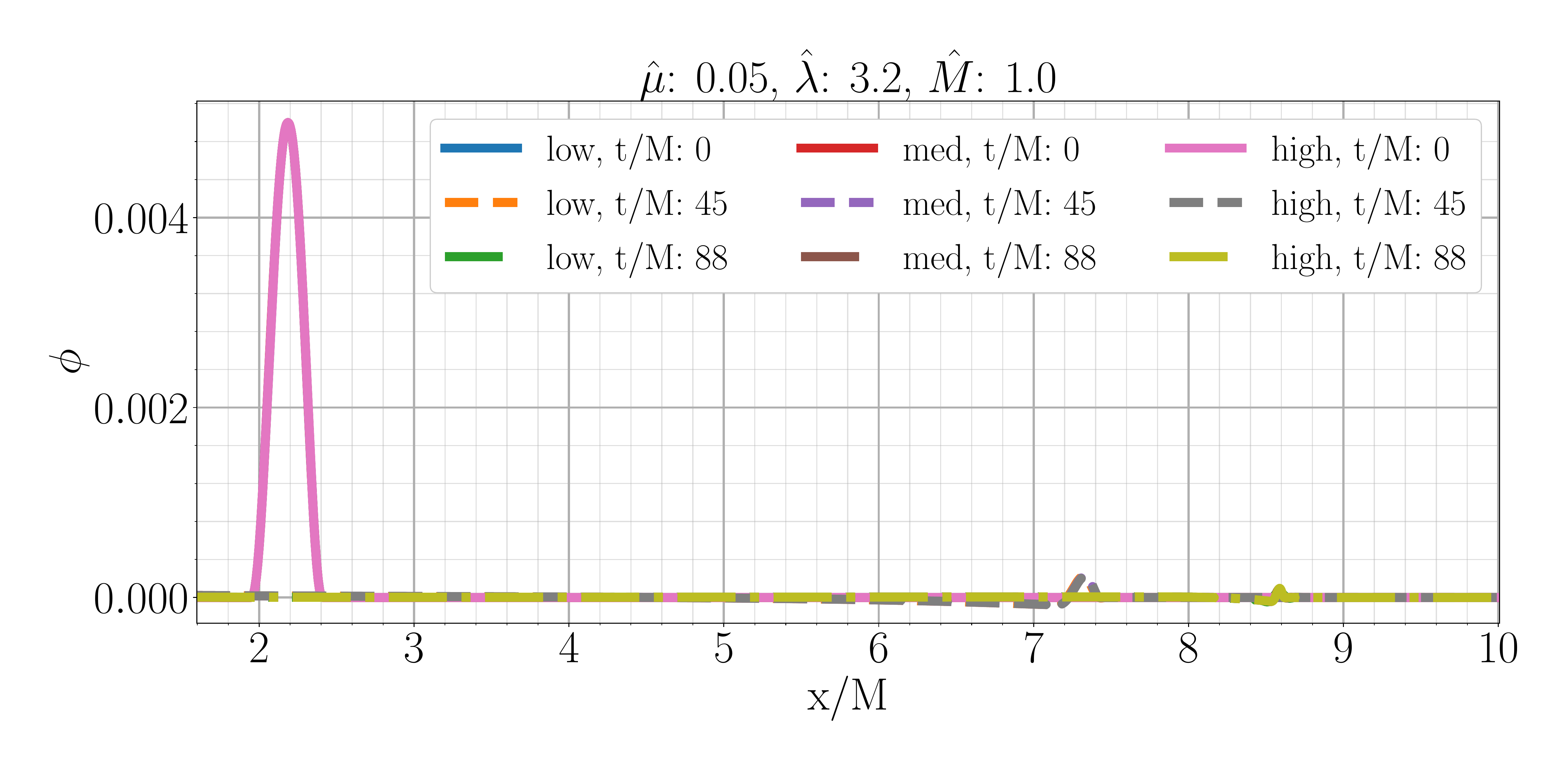}}}%
	\label{subfig:res_stidy_phi_disperses}
	\hfill
	\centering
	\subfloat[Scalarized black hole forms]{{\includegraphics[width=0.7\textwidth]{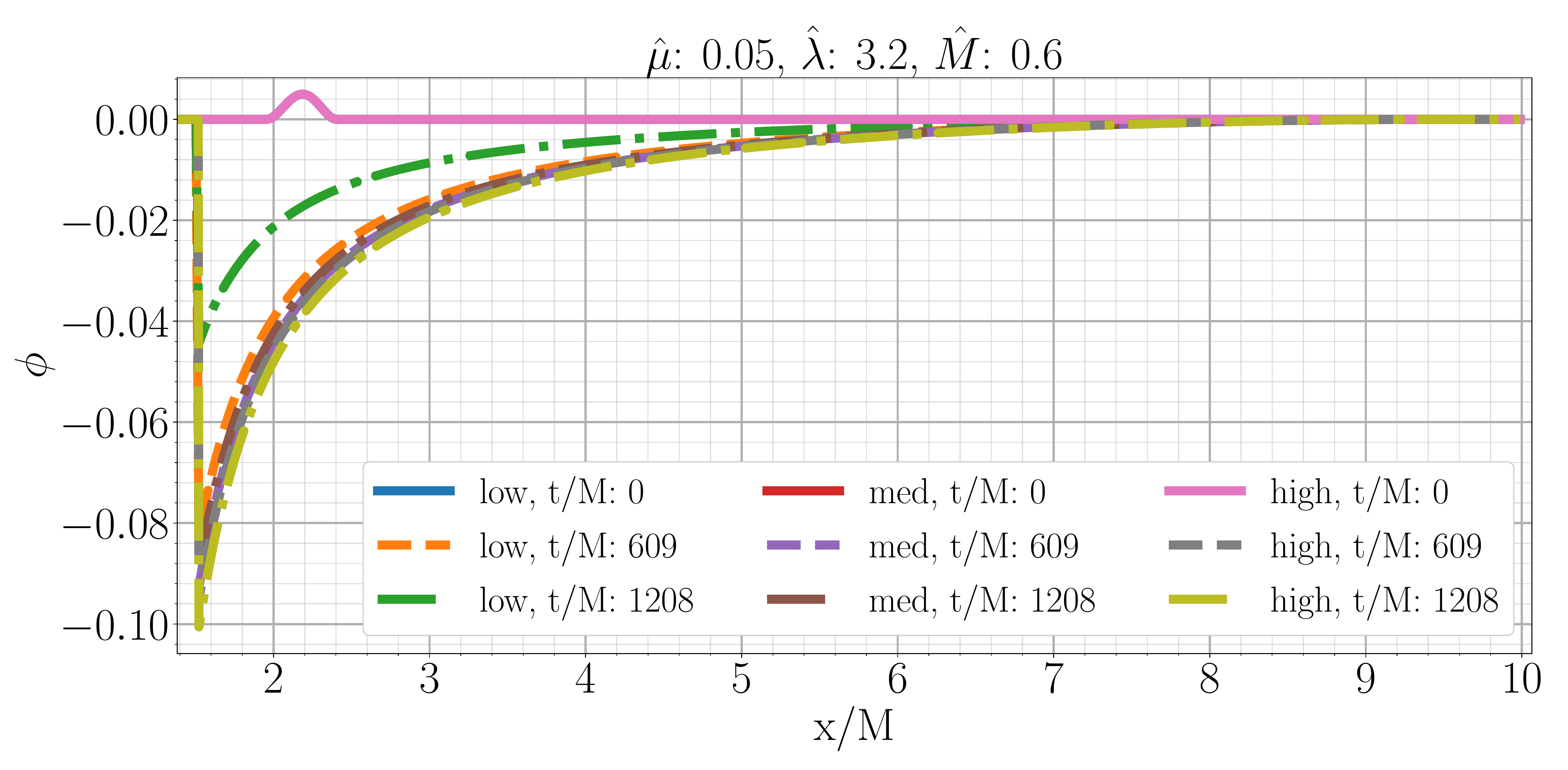}}}%
	\label{subfig:res_study_phi_scalarizes}
	\hfill
	\centering
	\subfloat[Elliptic region formation]{{\includegraphics[width=0.7\textwidth]{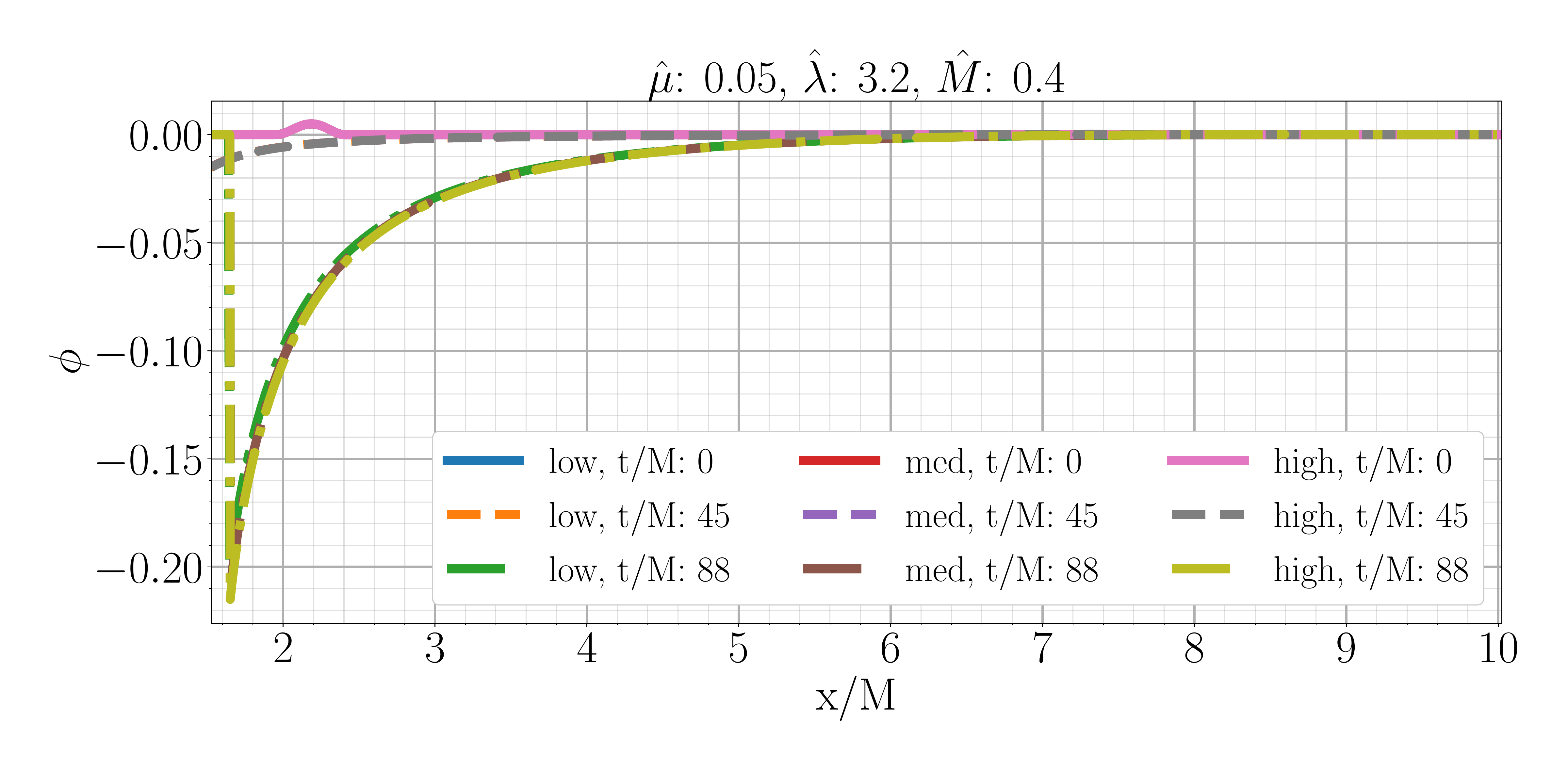}}}%
	\label{subfig:res_study_phi_ellp_form}
	\hfill
\caption{
Example evolution for EdGB solution with compact scalar field
initial data as described in Sec.~\ref{sec:compact_scalar_pulse_ID}.
Regarding the scalarized black hole case (b),
there is some truncation error induced decay of the scalar field at late
times, most evident in the lower resolution case (dash-dot green curve),
however with increasing resolution we do find convergence to a stable,
static scalarized black hole.
Low resolution for cases (a) and (c) 
is $N_x=2^{10}+1$ grid points, and for
case (b) $N_x=2^{9}+1$ grid points. In all cases
med and high resolution are 
double and quadruple the low resolution respectively.
}
\label{fig:res_studies_compact_scalar}
\end{figure*}

In Fig.~\ref{fig:res_studies_compact_scalar}
we show example scalar field solutions at three different
resolutions to demonstrate convergence for the three kinds of
behavior we generally observe in our simulations---decay to Schwarzschild,
scalarization, and development of an elliptic region.

\begin{figure*}
\centering
{\includegraphics[width=0.8\textwidth]{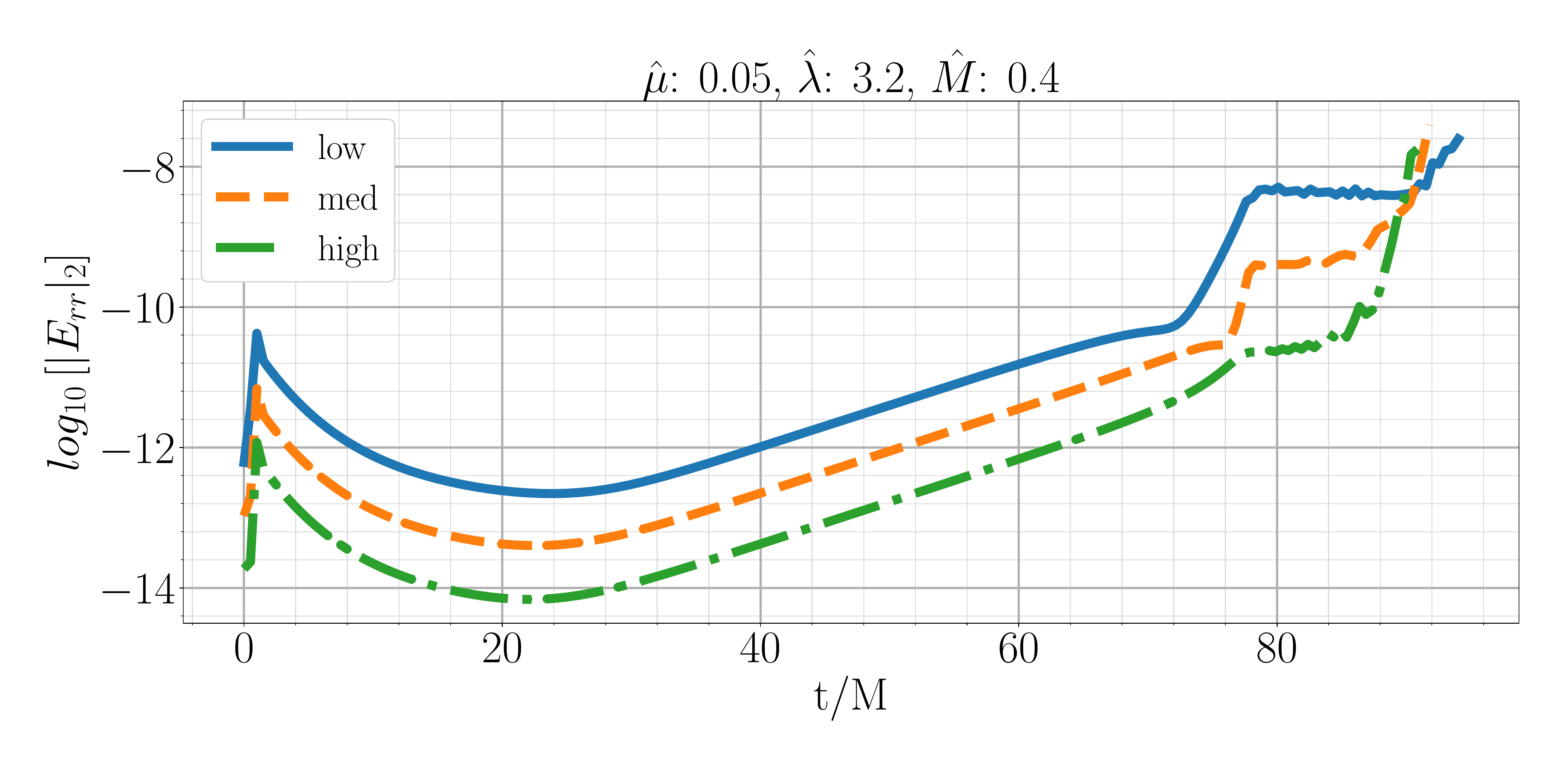}}%
\label{subfig:indep_res_ellp_form}
\caption{
Two-norm of the independent residual $E_{rr}$
for a run that forms a naked elliptic region, with compact scalar initial
data (see lower panel of Fig.~\ref{fig:res_studies_compact_scalar}).
We see convergence up until near
the formation of the elliptic region. Low resolution corresponds to
$N_x=2^{10}+1$ grid points, and med and high resolution correspond to
double and quadruple that number of grid points. 
See Sec. \ref{sec:compact_scalar_pulse_ID} for simulation parameters.
}
\label{fig:indep_res_ellp_formation}
\end{figure*}

	Fig.~\ref{fig:indep_res_ellp_formation} provides an example
of an independent residual for a run that formed an elliptic region,
demonstrating convergence of the solution prior to
the appearance of the elliptic region.
The mass of the initial scalar field for this case is
$M_{\phi}\approx 0.01$,
compared to the initial black hole mass of $M\approx10$. Once the scalar field
interacts with the black hole, it begins to grow near the horizon,
and an elliptic region grows and expands past the black hole horizon.
This behavior is qualitatively similar to what we found for
shift-symmetric EdGB gravity \cite{Ripley:2019aqj}.
\subsection{Numerical results:
	approximately scalarized black hole initial data
}
\label{sec:approx_scalarized_ID}
\begin{figure*}
	\centering
	\subfloat{{\includegraphics[width=0.7\textwidth]{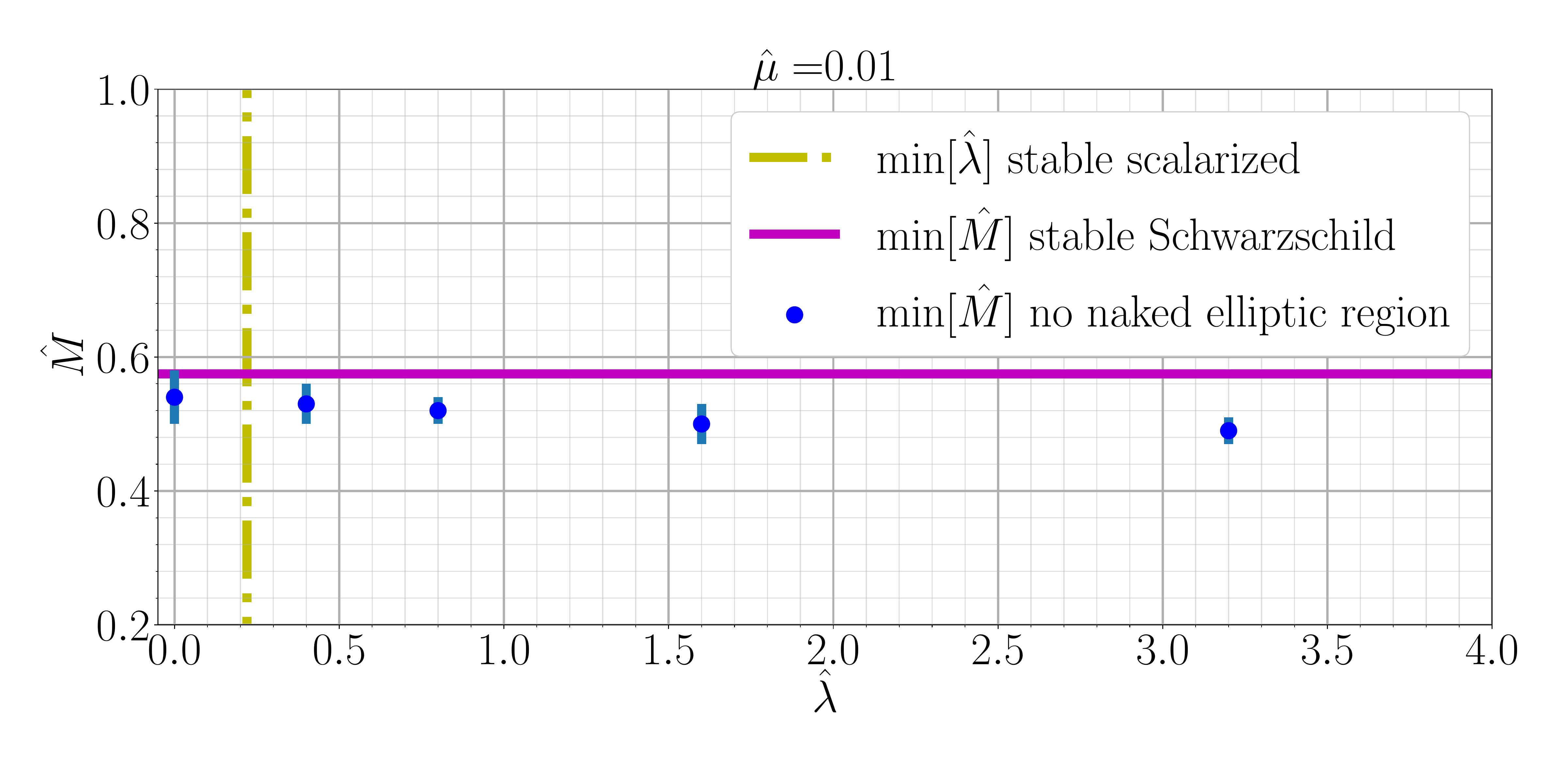}}}%
	\label{subfig:muhat0pt01_approx_scalarized}
	\hfill
	\centering
	\subfloat{{\includegraphics[width=0.7\textwidth]{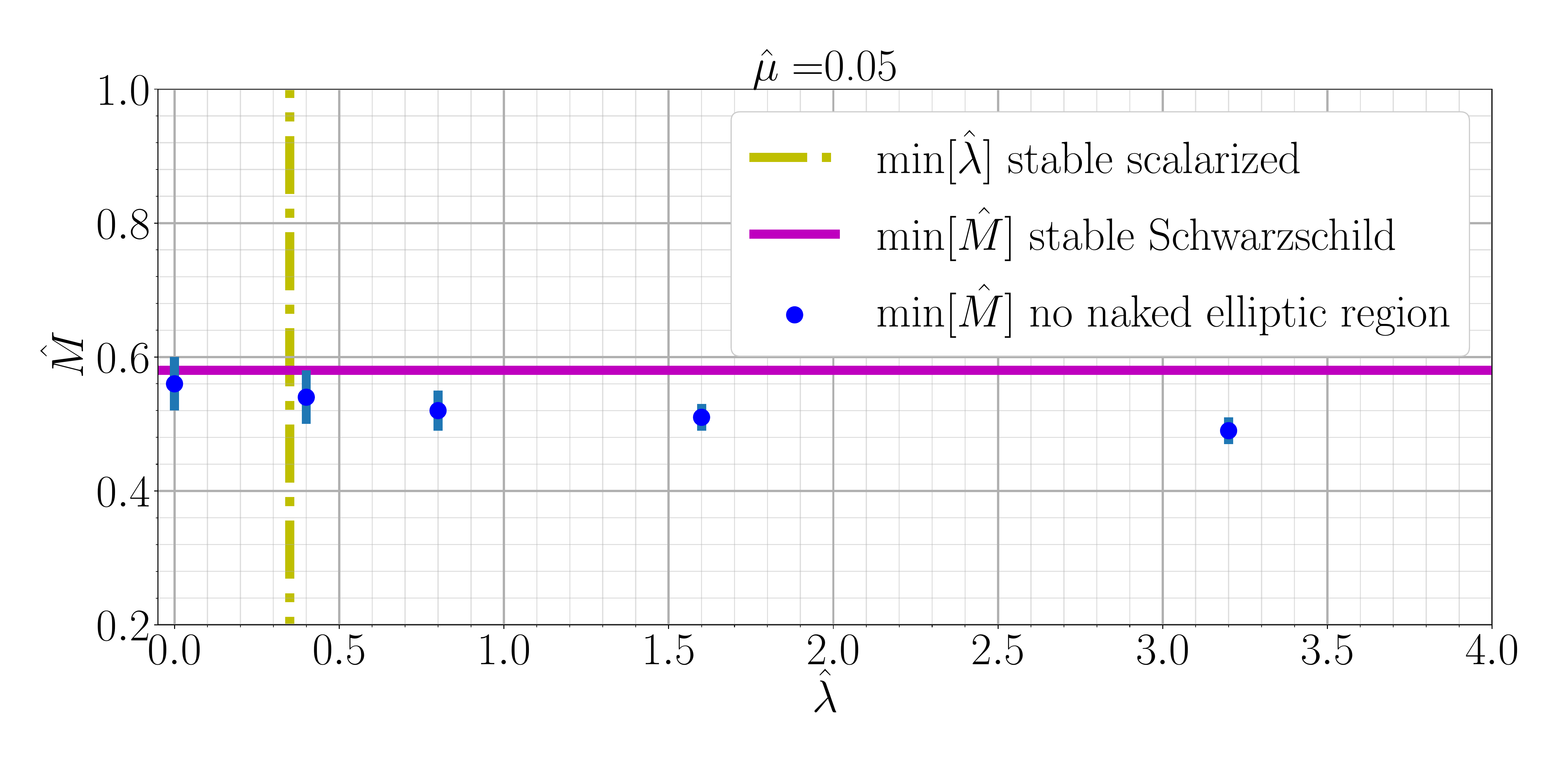}}}%
	\label{subfig:muhat0pt05_approx_scalarized}
	\hfill
	\centering
	\subfloat{{\includegraphics[width=0.7\textwidth]{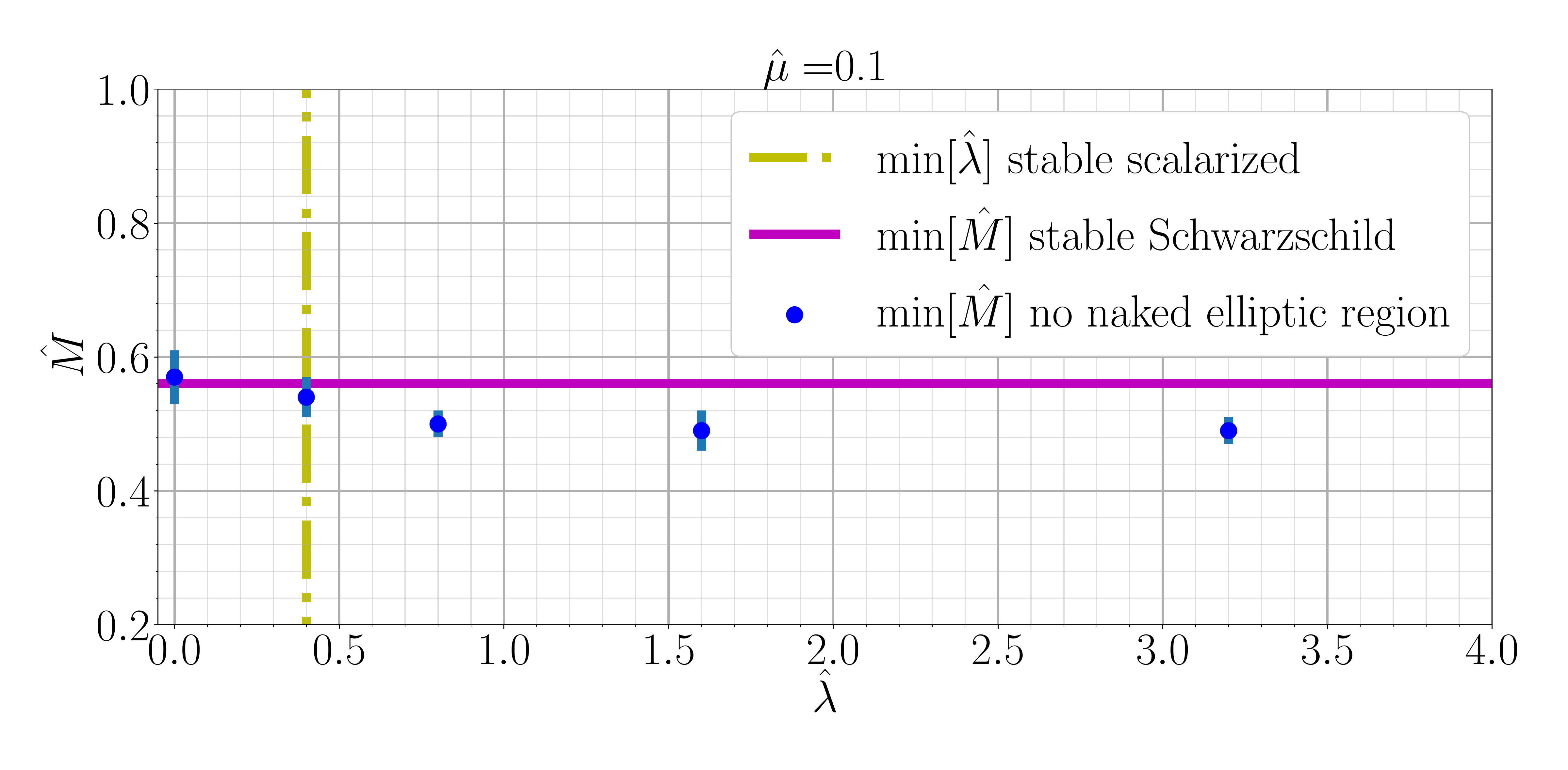}}}%
	\label{subfig:muhat0pt10_approx_scalarized}
	\hfill
\caption{
Onset of elliptic region formation, from evolution of approximate
scalarized black hole initial data as described in Sec.~\ref{sec:approx_scalarized_ID}.
This is the analog of Fig.~\ref{fig:compact_scalar}, and the same caption applies here.
For definitions of $\hat{M}$ and $\hat{\lambda}$ see respectively
Eq.~\ref{eq:hat_M} and Eq.~\ref{eq:hat_lambda}. 
}
\label{fig:approx_scalarized}
\end{figure*}

\begin{figure*}
\centering
\includegraphics[width=0.8\textwidth]{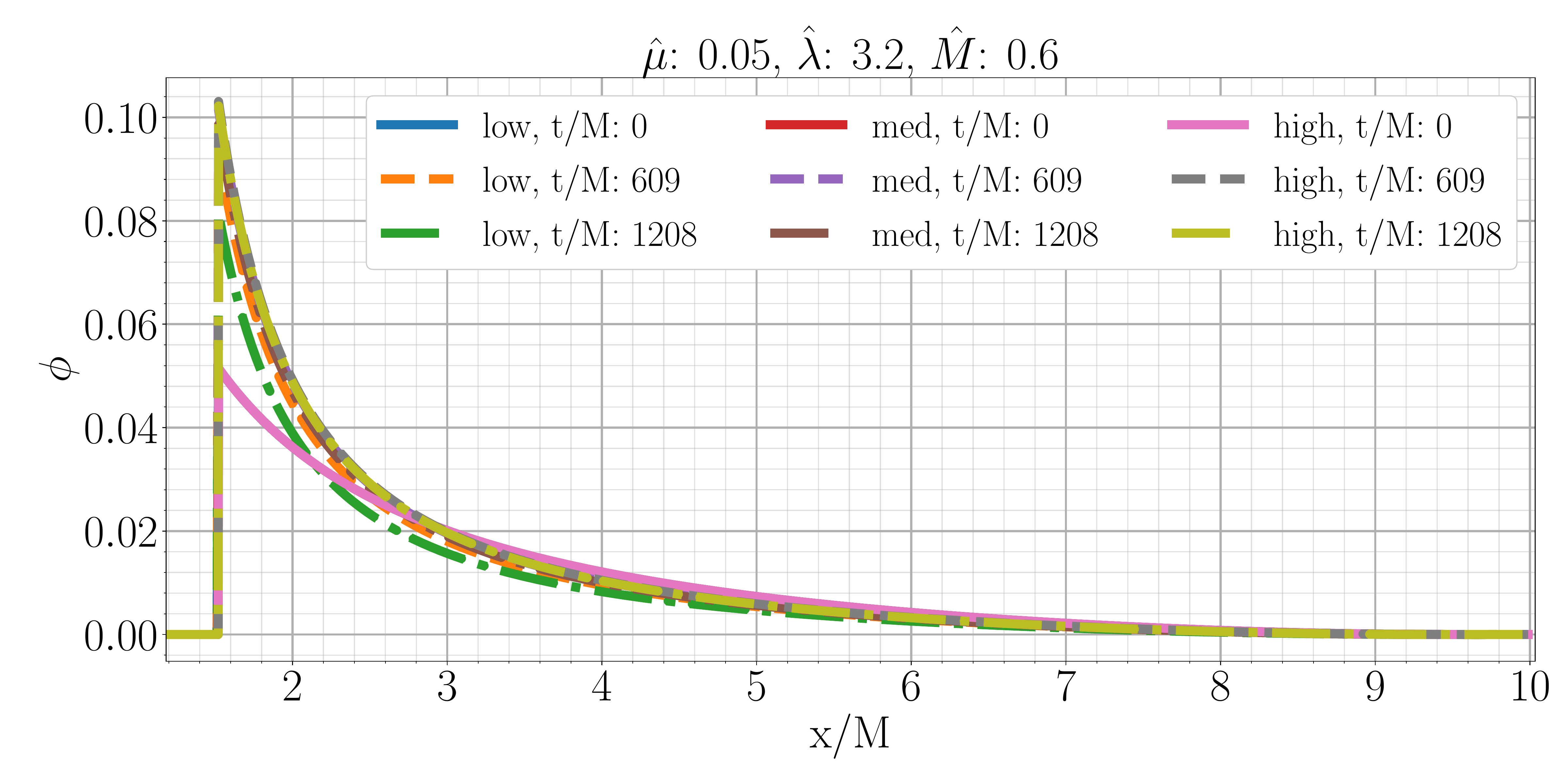}%
\hfill
\caption{
Formation of stable scalarized black hole from approximate
scalarized initial data as described in Sec.~\ref{sec:approx_scalarized_ID}.
We show runs with three different resolutions:
`low', `med', and `high'; `low' resolution corresponds to $N_x=2^{10}+1$
radial grid points and `med' and `high' correspond to double
and quadruple this number of radial points.
}
\label{fig:res_study_phi_scalarizes_scalarized}
\end{figure*}

The results of evolving the approximate scalarized initial data described
in Sec.~\ref{sec:initial_data_approx_scalarized} are qualitatively
similar to that of the perturbed Schwarzschild case described in
the previous section;  Fig.~\ref{fig:approx_scalarized} is
the analogous plot to that of Fig.\ref{fig:compact_scalar} to illustrate.
For the relevant initial data parameter we chose $\Phi_0=0.05$ and $M=10$
in Eq.~\ref{eq:scalarized_id}.
We found empirically that choosing $\Phi_0=0.05$
provided a scalarized profile reasonably close to the final stable scalarized
profiles for the $\hat{M}$, $\hat{\mu}$, and we considered
(for the definition of $\hat{M}$, $\hat{\mu}$, and $\hat{\lambda}$
see respectively Eq.~\ref{eq:hat_M}, Eq.~\ref{eq:hat_mu},
and Eq.~\ref{eq:hat_lambda}).
See Fig.~\ref{fig:res_study_phi_scalarizes_scalarized}
for an example of evolution of this initial data to a
stable scalarized profile. 

We also note that 
the onset of elliptic region formation in this parameter space does appear
to depend on the initial data as well. Comparing
Fig.~\ref{fig:compact_scalar} and
Fig.~\ref{fig:approx_scalarized} we see that our results support
the conclusion that an elliptic region may form at a larger value
of $\hat{M}$ when we start with approximately scalarized initial data. 

\section{Discussion}
\label{sec:discussion}
	We have numerically investigated perturbed black hole solutions in a 
$\mathbb{Z}_2$ symmetric
(i.e. a theory with an action invariant under the operation $\phi\to-\phi$)
variant of EdGB gravity.
We have found, consistent with the linear analysis of \cite{Macedo:2019sem},
that stable scalarized 
black holes exist within this theory.
However, for sufficiently large couplings relative to the scale of the black hole
the theory dynamically loses hyperbolicity: the scalar field
grows around the black hole until an elliptic region expands past
the black hole horizon. These results, along with the recent results
of \cite{Kovacs:2020ywu,Kovacs:2020pns}, suggests that in a limited
parameter range scalarized
black holes are subject to well-posed hyperbolic evolution. It would
be interesting to see whether this conclusion extends beyond spherical symmetry,
for example for binary black hole inspiral and merger.
The existence of naked elliptic regions in the theory for sufficiently
large couplings though strongly suggests that
the theory makes most sense from an effective field theory
point of view (which was the original motivation for the
particular form of EdGB gravity studied here\cite{Macedo:2019sem}).

The code\cite{justin_ripley_2020_3873503}
we wrote and used to produce the simulation results presented
here can easily be altered to accommodate other
forms of Gauss-Bonnet coupling $W(\phi)$ and
scalar field potential $V(\phi)$.
It would be interesting to investigate how these potential functions
influence the structure of scalarized black holes that can form,
and the region of solution space hampered by naked elliptic regions.
It would also be interesting to include the $(\nabla\phi)^4$ term in
action, to see the full range of dynamics that could occur for this
class of $\mathbb{Z}_2$ symmetric scalar-tensor theories.
\ack
J.L.R. thanks Nico Yunes and Helvi Witek for hospitality and useful
discussions at the University of Illinois, Urbana Champaign,
where some of the work for this project was accomplished.
F.P. acknowledges support from NSF
grant PHY-1912171, the Simons Foundation, and the
Canadian Institute For Advanced Research (CIFAR).
\appendix
\section{Characteristics versus linearized perturbation analysis}
\label{sec:comparison}
	Here we provide a brief discussion of the difference between
linear perturbation theory analysis to identify stable/unstable modes
(which is what is done in \cite{Macedo:2019sem})
and characteristic analysis (which is the analysis we
perform in our code).

	The scalar field offers the only dynamical degree of freedom
in EdGB gravity in spherical symmetry.
The authors in \cite{Macedo:2019sem} linearized the dynamics of that
degree of freedom (which we denote by $\varphi$),
and found that for a \emph{static} background
it obeys an equation of the form 
\begin{equation}
\label{eq:linearized_eq_Macedo}
	h(r)\frac{\partial^2\varphi}{\partial t^2}
-	\frac{\partial^2\varphi}{\partial r^2}
+	k(r) \frac{\partial\varphi}{\partial r}
+	p(r)\varphi
	=
	0
	,
\end{equation} 
	where $h,k,p$ are functions of the static background geometry.
They considered solutions of the form $\varphi(t,r)= e^{i\omega t}\psi(r)$, and
searched for conditions that would make $\omega^2<0$.

The characteristics of Eq.~\ref{eq:linearized_eq_Macedo} are found by keeping
only highest derivative terms and replacing
$\partial_t\to\xi_t$, $\partial_r\to\xi_r$, where $\xi_a\equiv(\xi_t,\xi_r)$
is the characteristic vector (see e.g.
\cite{lax1973hyperbolic,kreiss1989initial},
or\cite{Ripley:2019hxt,Ripley:2019irj}  in the context of
shift symmetric EdGB gravity),
and finding the zeros of this principal symbol of the system:
\begin{equation}
\label{eq:characteristic_eq_Macedo}
	h(r)\xi_t^2
-	\xi_r^2
	=
	0
	.
\end{equation} 
	The solutions $c\equiv-\xi_t/\xi_r$ to this equation give
the radial characteristic speeds.
Finding $\omega^2<0$ solutions
to Eq.~\ref{eq:linearized_eq_Macedo} can simply indicate a particular
background solution is unstable to perturbations.
In that case, as long as the theory remains hyperbolic the unstable
solution could evolve to a different, stable one.
By contrast, finding a $c^2<0$ solution to
Eq.~\ref{eq:characteristic_eq_Macedo} indicates a breakdown of
hyperbolicity of the theory evaluated at that solution.
When hyperbolicity has broken down, the equations of motion
can no longer be solved as evolution equations.

We emphasize that the form of the characteristic equation we solve,
Eq.~\ref{eq:characteristic_eqn},
does \emph{not} take the form Eq.~\ref{eq:characteristic_eq_Macedo},
as we solve the characteristic equation within a dynamical spacetime,
and we use a different coordinate system than is used in \cite{Macedo:2019sem}.
Instead the characteristic equation takes the schematic form
(compare to Eq.~\ref{eq:cartoon_characteristics})
\begin{equation}
	\mathcal{A}\xi_t^2-\mathcal{B}\xi_t\xi_r+\mathcal{C}\xi_r^2
	=
	0
	,
\end{equation} 
	where $\mathcal{A}$, $\mathcal{B}$, and $\mathcal{C}$
depend on the background fields
$\alpha$, $\zeta$, $P$, $Q$, and their radial and time derivatives.

From the forms of
Eq.~\ref{eq:linearized_eq_Macedo} and
Eq.~\ref{eq:characteristic_eq_Macedo} it is clear that linear
(in)stability does not necessarily imply (lack of) hyperbolicity, or
vice-versa.
Nevertheless our hyperbolicity analysis does conform with the general
results of \cite{Macedo:2019sem}.
\section{Approximate scalarized profile}
\label{sec:approx_scalarized_profile}
	Here we provide a derivation of the approximate scalarized
black hole initial data presented in
Sec.~\ref{sec:initial_data_approx_scalarized}.

Consider a Schwarzschild background:
we set $\alpha=1$, $\zeta=\sqrt{2m/r}$,
set $\partial_t\phi=0$, and solve Eq.~\ref{eq:scalar_eom}
\begin{equation}
	\nabla_{\mu}\nabla^{\mu}\phi
-	\left(\mu^2-\frac{1}{4}\eta\mathcal{G}\right)\phi
-	4\lambda\phi^3
	=
	0
	.
\end{equation}
	Plugging things in, we find 
\begin{equation}
	\left(1-\frac{2m}{r}\right)\frac{d^2\phi}{dr^2}
+	2\left(1-\frac{m}{r}\right)\frac{1}{r}\frac{d\phi}{dr}
-	\left(\mu^2-\frac{12\eta m^2}{r^6}\right)\phi
-	4\lambda\phi^3
	=
	0
	.
\end{equation} 
	In the far field limit $(r\to\infty)$, and assuming
$\lim_{r\to\infty}\phi=0$, we have to leading order
\begin{equation}
	\frac{d^2\phi}{dr^2}
+	\frac{2}{r}\frac{d\phi}{dr}
	=
	\mu^2\phi
	.
\end{equation}
	The solution consistent with the field going
to zero at spatial infinity is
\begin{equation}
	\phi(r)
	=
	\frac{C}{r}\mathrm{exp}\left(-\mu r\right)
	,
\end{equation}
	where $C$ is a constant. We can split up this constant into two
as follows:
\begin{equation}
	\phi(r)
	=
	\frac{c_1}{r}\mathrm{exp}\left(-\mu (r-c_2)\right)
	.
\end{equation}
	In our simulations we set $c_2=3m$ on the initial data slice,
see Eq.~\ref{eq:scalarized_id}.
\section*{References}
\bibliography{localbib}
\end{document}